\documentclass[aps,pra,reprint,superscriptaddress,nobibnotes]{revtex4-2}

\usepackage{graphicx}% Include figure files
\usepackage{dcolumn}% \documentclass[aps,prl,reprint,superscriptaddress,nobibnotes]{revtex4-2}
\usepackage{blindtext}
\usepackage{lineno}
\usepackage{amsmath}
\usepackage{epstopdf}
\usepackage{float}
\usepackage{txfonts}
\usepackage[utf8]{inputenc}
\usepackage[T1]{fontenc}
\usepackage{subfigure}

\usepackage{esint}
\usepackage{braket}
\usepackage{hyperref}

\usepackage[separate-uncertainty=true]{siunitx}

\usepackage[dvipsnames]{xcolor}

\colorlet{mylinkcolor}{RoyalPurple}
\colorlet{mycitecolor}{RoyalPurple}
\colorlet{myurlcolor}{RoyalPurple}
\usepackage{hyperref}
\hypersetup{
	linkcolor  = mylinkcolor,
	citecolor  = mycitecolor,
	urlcolor   = myurlcolor,
	colorlinks = true,
	breaklinks = true
}

\DeclareSIUnit{\rad}{rad}
\DeclareSIUnit{\px}{px}

\newcommand{\subfig}[1]{(#1)}

\begin{document}
\title{Hybrid quantum memory leveraging slow-light and gradient-echo duality}
\author{Stanisław Kurzyna}
\email{s.kurzyna@cent.uw.edu.pl}

\affiliation{Centre for Quantum Optical Technologies, Centre of New Technologies, University of Warsaw, Banacha 2c, 02-097 Warsaw, Poland}
\affiliation{Faculty of Physics, University of Warsaw, Pasteura 5, 02-093 Warsaw, Poland}
\author{Mateusz Mazelanik}
\affiliation{Centre for Quantum Optical Technologies, Centre of New Technologies, University of Warsaw, Banacha 2c, 02-097 Warsaw, Poland}
\author{Wojciech Wasilewski}
\affiliation{Centre for Quantum Optical Technologies, Centre of New Technologies, University of Warsaw, Banacha 2c, 02-097 Warsaw, Poland}
\affiliation{Faculty of Physics, University of Warsaw, Pasteura 5, 02-093 Warsaw, Poland}
\author{Michał Parniak}
\email{michal.parniak@uw.edu.pl}
\affiliation{Centre for Quantum Optical Technologies, Centre of New Technologies, University of Warsaw, Banacha 2c, 02-097 Warsaw, Poland}
\affiliation{Faculty of Physics, University of Warsaw, Pasteura 5, 02-093 Warsaw, Poland}

\begin{abstract}
We demonstrate a hybrid quantum memory that combines Gradient Echo Memory (GEM) and Electromagnetically Induced Transparency (EIT) protocols for reversible mapping between light and atomic coherence. By leveraging GEM and EIT complementarity, we realize time-to-frequency and frequency-to-time conversion mechanisms for spectro-temporal modes. This capability provides a versatile tool for quantum communication, where coherent frequency–time conversion enhances network interoperability. In addition, the protocol may enable fundamental studies of atomic coherence, including investigations of Rydberg polaritons and mapping of single Rydberg excitations and ionic impurities.
\end{abstract}
\maketitle

\section{Introduction}

Since their introduction as building blocks of quantum repeaters \cite{Duan2001,Lvovsky2009}, quantum memories have been proposed and implemented across a wide range of physical systems \cite{heshami_quantum_2016}. The diversity of platforms has significantly broadened their functionality beyond the original objective of enabling quantum networks \cite{Kaneda2017,Humphreys2014,Niewelt2023,Brecht2015,Viscor2012,Chang2014}. In particular, multimode memories with extended coherence times substantially improve the performance of quantum repeaters \cite{Bao2012,Li2021,Albrecht2015,Zhang2024} and, when endowed with intrinsic processing capabilities, can function as transducers between distinct domains \cite{Vashukevich2020,Fisher2016}, including time–frequency \cite{Mazelanik:20} and frequency–position \cite{Jastrzebski2024}. Additionally, the capability to store optical photons while enabling controlled manipulations has facilitated the realisation of quantum-enhanced sensing and metrology protocols that implement optimal measurement on the optical state \cite{Mazelanik2022}.

The control in such protocols implemented in atomic memories is achieved through the modulation of atomic coherence established during the write-in stage. The resulting spatial distribution of this coherence encodes the spatial and spectro-temporal properties of the incident optical signal. Specifically, the Gradient Echo Memory (GEM) protocol \cite{Sparkes2013, Higginbottom2012} realises a spectrum-to-position mapping by introducing inhomogeneous broadening of the medium’s absorption profile via a magnetic-field gradient or electric-field gradient \cite{Alexander2006,Sparkes2010,hetet_electro-optic_2008}.
In contrast, the Electromagnetically Induced Transparency (EIT) protocol \cite{Heinze2013,Ma2017} relies on strong dispersion within the medium \cite{Kang2003,sautenkov_ultradispersive_2010,Mikaeili2022}, whereby the optical pulse is substantially slowed during propagation through the ensemble and subsequently halted after a controlled delay. This process maps the temporal profile of the input field onto the atomic coherence along the propagation direction, thereby implementing a time-to-position mapping \cite{Hsiao2018,Chen2013}. These two approaches are therefore complementary.
It is important to notice that neither time-to-frequency nor frequency-to-time mappings can be obtained solely by either of GEM or EIT operations without additional modification of the protocol. Writing in GEM allows for mapping the frequency profile to the position and readout in GEM maps it back to the temporal profile of the input pulse reversed in time \cite{hetet_multimodal_2008}. 

Another way to implement frequency-to-position mapping with the multi-mode capabilities is realised in atomic frequency combs (AFC) memories \cite{Afzelius2009,Riedmatten2008,Sinclair2014,Ma2021}. Due to the multimodality of AFC memories, photons with different arrival times can be stored in the memory, allowing them to be used as time-bin qubits. However, AFCs are mainly created with the warm atomic vapours or rare-earth-doped crystals, which allows for the large memory bandwidth but makes them incompatible for narrowband photons as compared to GEM or EIT. 
Moreover, contrary to the GEM spectrum-to-position mapping, in the AFC memories based on crystals, the positions of the spectral components are not well defined. Furthermore, the AFC protocol in warm atomic vapours maps the spectral components to different velocity classes. 

Importantly, both EIT and GEM mappings are reversible, enabling direct investigation of the internal structure of atomic coherence through the detection of emitted light during the read-out stage. This is particularly relevant, as coherence may undergo unknown phase and amplitude modulations induced by external fields or interactions with ionic or Rydberg impurities \cite{Du:25,Gross2020,Guenter2013}.

Recently, the combination of those two approaches was used to perform the frequency to time mapping and Fourier imaging \cite{Papneja:26}. This demostrated the time-frequency Fourier transform performed on the light stored in the memory, as well as the intra-memory interference realized by combining frequency-to-position mapping in GEM and position-to-time in EIT.
In this work, we demonstrate a hybrid quantum memory that integrates the GEM and EIT protocols to perform write-in and read-out of the atomic coherence, enabling the mapping of its $z$-component (along the propagation direction) onto either frequency or time of the retrieved optical field. By implementing the two protocols in complementary configurations, we show time-to-frequency and frequency-to-time conversions. Using the heterodyne detection scheme for the read-out light, we were able to investigate the mapping simultaneously in the time and frequency domains.

\section{Theory}

\begin{figure*}[t]
\centering
\includegraphics[width=2\columnwidth]{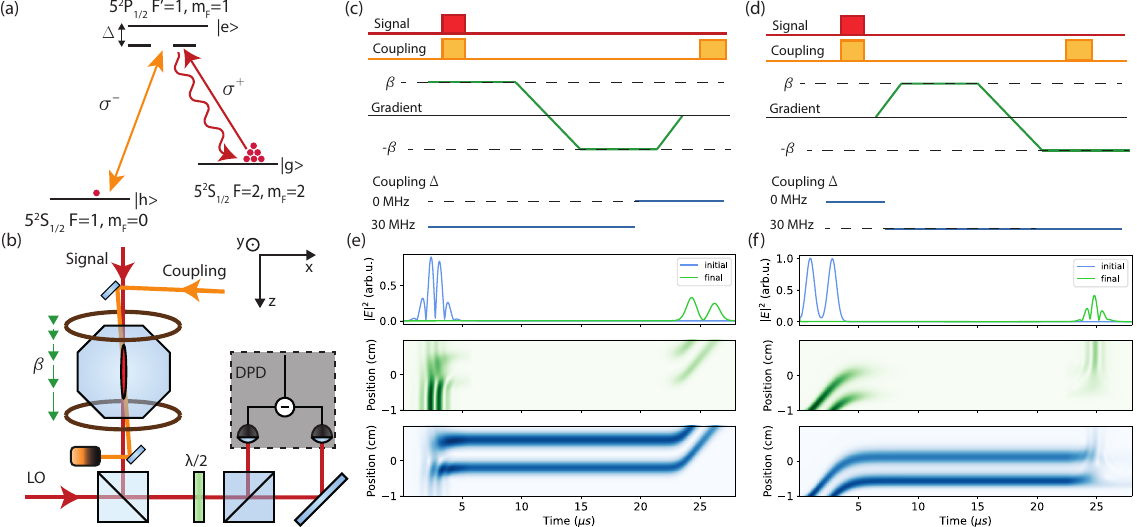}
\caption{
\subfig{a} Energy levels of the $^{87}$Rb relevant for the experiment.
\subfig{b} Experimental setup for the presented protocol. A differential photodiode (DPD) is used to detect the beating between the signal and the local oscillator (LO).
\subfig{c} Experimental sequence for storage in the GEM protocol and readout in the EIT regime.
\subfig{d} Experimental sequence for stopping light under the EIT regime and readout in the GEM protocol.
\subfig{e,f} Numerical simulations for the storage in GEM and readout in EIT, and stopping the light in EIT and readout in GEM, respectively. The upper row demonstrates the time trace of the detected pulse. The middle row represents the intensity of the electrical field of the write-in and readout. The lower row represents the absolute value of the atomic coherence induced between states $\ket{g}$ and $\ket{h}$.  
}
\label{fig:slow_GEM_schemat}
\end{figure*}

Let us consider a $\Lambda$ type photon-atom interface depicted in Fig.~\ref{fig:slow_GEM_schemat}\subfig{a} in which GEM and EIT protocols can be implemented. The corresponding experimental schematic is displayed in Fig.~\ref{fig:slow_GEM_schemat}\subfig{b}, which also includes gradient coils for the GEM protocol.

In GEM, during the write-in and read-out stages, atoms are placed in a magnetic field gradient. This induces inhomogeneous broadening of the atomic medium due to Zeeman shifts of the magnetically sensitive storage states $|g\rangle$ and $|h\rangle$. The frequency shift, or detuning, can be written as:
\begin{equation}\label{eq:gem}
    \delta(z) = \beta \cdot z +\omega_0,
\end{equation}
where $\beta$ is value of the magnetic gradient, $z$ is the position on the atomic ensemble and $\omega_0$ is the resonance frequency of the two-photon transition.
Thanks to that, the frequency components of the input optical field (signal) are stored in the different parts of the atomic cloud according to the formula:
\begin{equation}
    \varrho_{gh}(T,z)\propto e^{i\beta z T}\tilde{A}(\beta z),
\end{equation}
where $\varrho_{gh}$ is the atomic coherence between states $|g\rangle$ and $|h\rangle$, $\tilde{A}(\omega)$ is a Fourier transform of the signal's temporal shape $A(t)$ and $T$ is the duration since arrival of the input pulse. 
During the storage in the magnetic field gradient, the coherence acquires the additional wavevector $k_z = \beta T$, which allows for mapping times of arrival to wavevectors. This provides the possibility to map the temporal profile to the wavevector domain and allows for storing multiple subsequent impulses and reading them in reverse order. 
To read out the stored coherence, it is necessary to unwind the magnetic-field-induced phase $e^{i\beta z T}$ acquired during the storage. This is performed by reversing the gradient during the read-out stage. Due to reversible mapping, the read-out electric field is proportional to the Fourier transform of the atomic coherence. Therefore, the final electric field is proportional to the initial pulse but mirrored in time.        
% \paragraph*{EIT}

A complementary mapping, i.e., in which the temporal profile of the signal light is mapped to the position in the medium, can be achieved using EIT protocol \cite{Phillips2001}. Due to the extremely high electric susceptibility under the two-photon resonance, the refractive index of the atomic medium increases accordingly. 
This leads to the vast decrease of the group velocity of the pulse propagating through the atomic medium, causing it to travel as the so-called slow-light polariton \cite{fleischhauer_quantum_2002}, which is spatially compressed.
Moreover, if the coupling beam is switched off, the group velocity of the propagating pulse is reduced to 0, allowing the signal pulse to be brought to a halt in the atomic memory. The group velocity follows the formula:
\begin{equation}
    v_g = \frac{c}{1+\frac{\mathrm{OD} \, \Gamma c g }{L\, \hbar \, \Omega_C^2}}
\end{equation}
where $g$ is atom-field coupling constant, $\Gamma$ is decay rate of the excited state, $c$ is the speed of light in vacuum, $\mathrm{OD}$ is the optical depth and $\Omega_C$ is the Rabi frequency of the coupling beam. 
Contrary to the GEM protocol, photons arriving at different times are stored in different parts of the atomic medium, mapping the temporal profile of the pulse to the position in the memory medium. The resulting coherence when neglecting dispersion can be written as:
\begin{equation}
    \varrho_{gh}(T,z)\propto A\left(\frac{v_gT-z}{v_g}\right).
\end{equation}
When the coupling beam is switched on again after the storage, the created stopped polariton begins to propagate again through the memory. 
The final electric field after the propagation is proportional to the initial pulse.  

By combining those two procedures, i.e., storing the light in the GEM protocol and reading the stored atomic coherence under the EIT condition, one can perform the frequency-to-time mapping. 
Storing the light in GEM allows mapping of the frequency of the incoming pulse to the position in the atomic memory. To read out the light from the memory with the highest efficiency, the coherence's central wavevector $k_z=\beta T$ acquired due to the magnetic field has to be reverted to zero by unwinding the magnetically induced phase with an additional gradient impulse with opposite value. 
However, due to the spread of the wavevectors of the stored coherence, it is impossible to remove the induced magnetic part entirely; hence, only the central wavevector is shifted to zero.
The efficient readout of the light stored in the GEM in the EIT regime can be achieved when the EIT bandwidth $\Delta \omega_{EIT} = \frac{\Omega_c^2 L}{\mathrm{OD} \, \Gamma }$ is greater than the frequency shift of the stored impulse, due to the aqcuiring the additional position-dependent GEM phase. We define the frequency shift as:
\begin{equation}
    \delta \omega = \frac{\partial \varphi}{\partial t} = \frac{\partial \varphi}{\partial z}\frac{\partial z}{\partial t} = \beta T v_g
\end{equation}
From this assumption, we get the condition for the efficient readout $T \beta L \leq \sqrt{OD}$. 

The coherence mapping using the combination of those two protocols can also be performed oppositely, i.e., storing the light by stopping the pulse in the EIT and reading it out in the GEM. 
Stopping the light in the EIT maps the time of the arrival of the incoming pulse to the position on the atomic ensemble. To apply the GEM protocol and read out the coherence with position mapped to the frequency, first, after propagation time in the EIT, the magnetic field gradient needs to be switched on. Then, the gradient needs to be reversed, and the coherence can be read out.

\section{Simulation}

\begin{figure*}
\centering
\includegraphics[width=2\columnwidth]{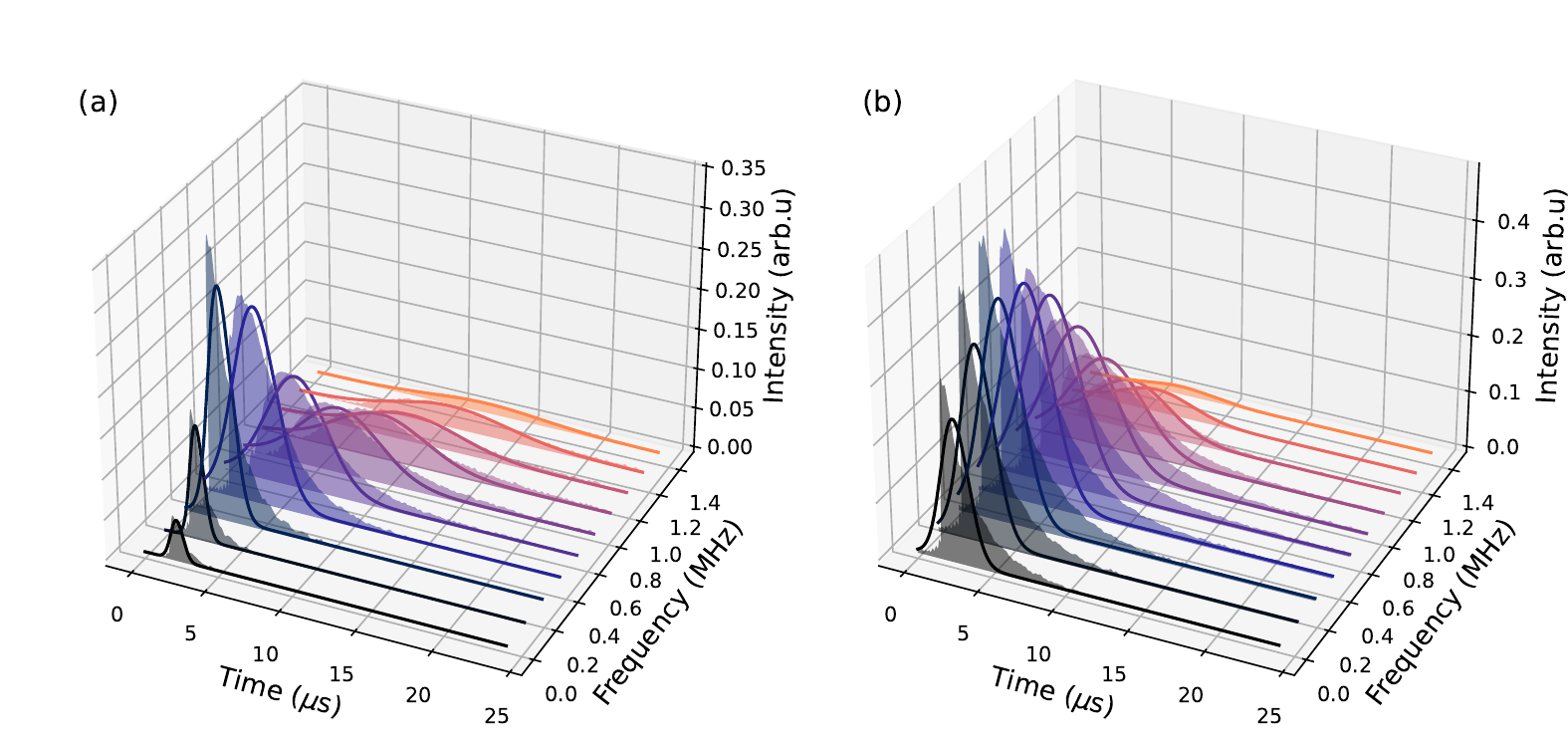}
\caption{\subfig{a} 
Spectrally narrow pulses stored in GEM and readout in EIT. Pulses are delayed according to the position of the mapped coherence after the GEM storage.
\subfig{b} Spectrally broad pulse stored in GEM and read out in EIT. The spectrum of the pulse is broader than the bandwidth of the memory, and the induced coherence covers the entire atomic cloud; thus, pulses are almost not delayed. The solid lines show a fitted Gaussian envelope.
}
\label{fig:frequency_delay25_3D}
\end{figure*}

\begin{figure*}
\centering
\subfigure{\includegraphics[width=1\columnwidth]{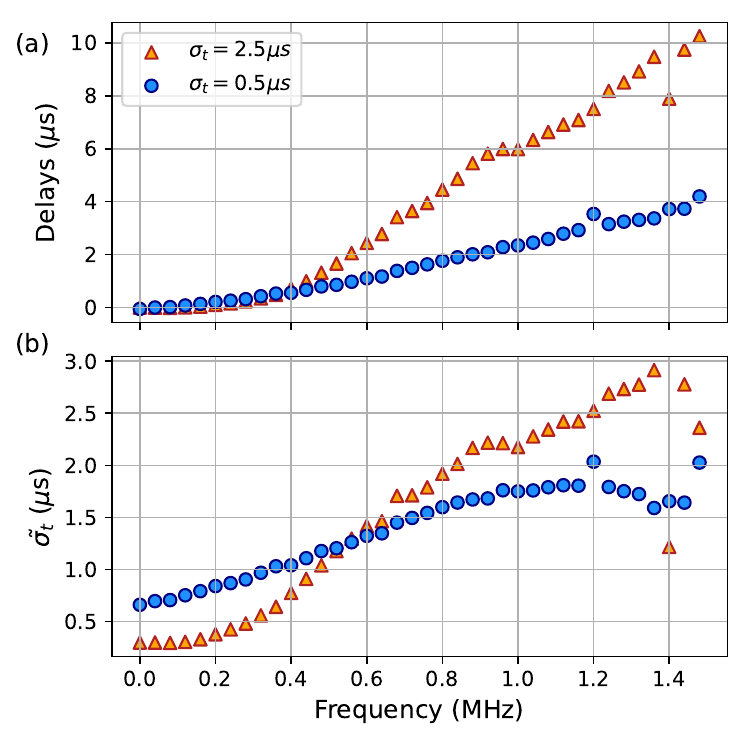}}
\subfigure{\includegraphics[width=1\columnwidth]{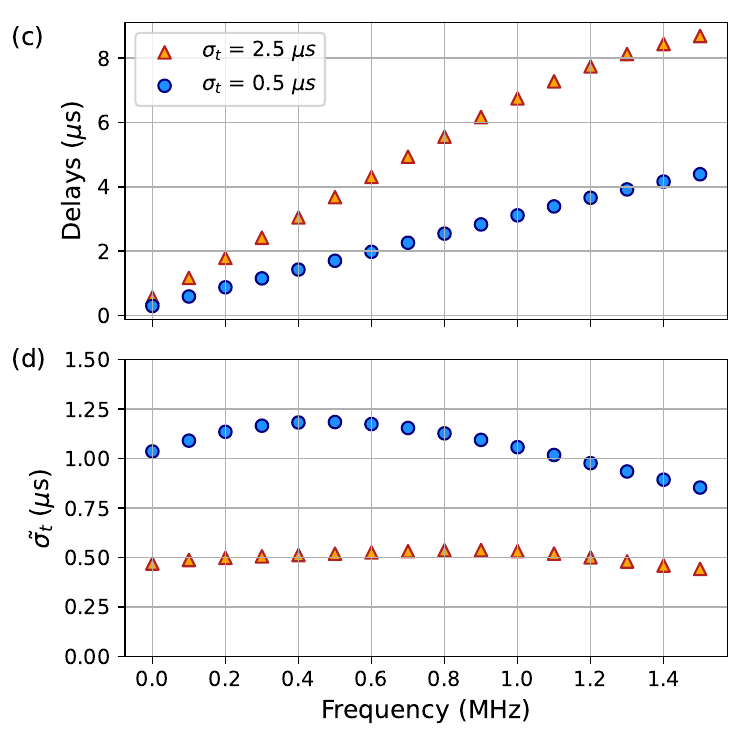}}
\caption{\subfig{a} Comparison of the delays for pulses with narrow and wide spectra, stored in GEM and read in EIT. Orange triangles correspond to the pulse with width $\sigma = \SI{2.5}{\micro \s}$ and blue dots represent the pulse with width $\sigma = \SI{0.5}{\micro \s}$. It is important to notice that the spectrally narrow pulse is significantly more delayed.
\subfig{b} The width of the fitted Gaussian function. 
\subfig{c} Comparison of the delays for the pulses with narrow and wide spectra obtained from the numerical simulations of the propagation of the pulses. 
\subfig{d} The widths of the fitted Gaussian function for the numerical simulations}
\label{fig:sim_EIT_freq_timeshifts}
\end{figure*}

Numerical simulations are performed by solving the optical Bloch equations for the propagating optical pulse in the GEM and for the pulse propagating under the EIT conditions. Without the loss of generality, we can show the procedure for the storage in the GEM and readout under EIT. The numerical simulation is performed using the XMDS2 package \cite{Dennis2013}. 
First, to find the electric field and atomic coherence in the GEM protocol by solving the set of propagation equations with one spatial dimension:
\begin{equation}
\begin{cases}
    \frac{\partial A}{\partial z} =  i \frac{n(z) \text{OD} \, \Gamma}{4\Delta + 2 i \Gamma}(\Omega_C \rho_{gh} + A) \\
    \frac{\partial \rho_{gh}}{\partial t} = \frac{i \Omega_C A}{4\Delta - 2 i \Gamma} - \rho_{gh}\left(\frac{i |\Omega_C|^2 }{4\Delta + 2 i \Gamma}  - i \delta (z) \right) 
\end{cases}    
\end{equation}
where $n(z)$ is spatial profile of the atomic memory and $\delta(z)$ is position dependent detuning. In our case, $\delta$ corresponds to the spatial magnetic gradient i.e., $\delta(z) = \beta z$. 
We chose the spatial atomic concentration to have a super-Gaussian shape with the width matching the experimentally measured length of the atomic cloud. 
The resulting atomic coherence and electric field obtained from the storage simulation in the GEM were used as the initial parameters for the simulation in the EIT regime.
To find the final electric field, we solve the set of equations corresponding to the propagation of the slow light polariton through the atomic medium:
\begin{equation}
    \begin{cases}
        \frac{\partial A}{\partial t} = -c \frac{\partial A}{\partial z} + i \frac{g_P}{2}P \\
        \frac{\partial P}{\partial t} = \frac{-\Gamma}{2}P + i \frac{g_P}{2}A + i \frac{\Omega_C}{2}S\\
        \frac{\partial S}{\partial t} = i \frac{\Omega_C}{2}P
    \end{cases}
\end{equation}
where $g_P = \sqrt{c \text{OD} \, \Gamma / L}$, $P = g_P \rho_{ge}$ and $S = g_P \rho_{gh}$. The results of the simulation for storing the pulse with two frequencies in GEM and reading in EIT as well as the reversed situation, i.e., storing two pulses in EIT and reading in GEM, are depicted in Fig.~\ref{fig:slow_GEM_schemat}\subfig{e} and \subfig{f}.

\section{Experimental setup}

\begin{figure*}
\centering
\includegraphics[width=2\columnwidth]{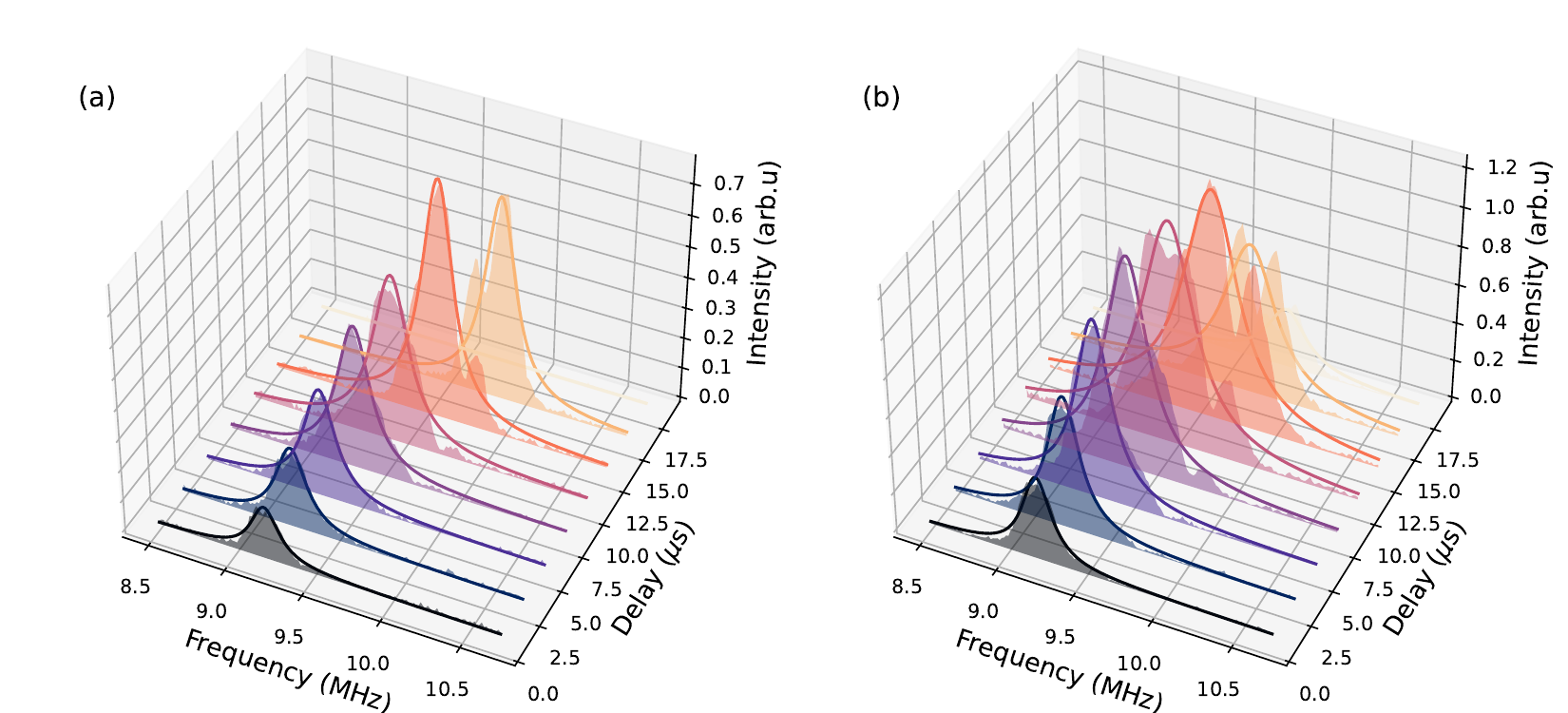}
\caption{\subfig{a} Fourier transform of the pulse, narrow in the time domain (i.e. $\sigma_t = \SI{0.5}{\micro \s}$), stored in the EIT and readout in the GEM. The readout frequency changes according to the position on the cloud where the pulse has stopped during the propagation.
\subfig{b} Fourier transform of the pulse, which is broad in the time domain (i.e. $\sigma_t = \SI{2.5}{\micro \s}$) stored in the EIT and readout in the GEM. The solid lines show a fitted Gaussian envelope.}
\label{fig:GEM_eit_delay05_3D}
\end{figure*}

The experiment is based on GEM, which is built on rubidium-87 atoms trapped in a magneto-optical trap (MOT). The trapping and experiments are performed in a sequence lasting \SI{12}{\ms}, resulting in $\SI{50}{\hertz}$ repetition rate. The sequences used in the experiment are presented in Fig.~\ref{fig:slow_GEM_schemat}\subfig{a,b}. Atoms form an elongated, cigar-shaped cloud with optical depth (OD) reaching 80 on the relevant transition. The ensemble temperature is \SI{80}{\micro\kelvin}. After the cooling and trapping procedures, atoms are optically pumped to the state $\ket{g} \coloneqq 5^2S_{1/2}\,F = 2, m_F = 2$. We utilize the $\Lambda$ system depicted in Fig.~\ref{fig:slow_GEM_schemat}\subfig{c} to couple the light and atomic coherence. Signal laser with $\sigma^{-}$ polarization is red detuned by $2 \pi \times \SI{30}{\mega\hertz}$ from the $\ket{g} \rightarrow \ket{e} \coloneqq 5^2P_{1/2}\,F = 1, m_F = 1$ transition. The coupling laser with $\sigma^{+}$ polarization is tuned to the resonance with the two-photon transition to state $\ket{h} \coloneqq 5^2S_{1/2}\,F = 1, m_F = 0$, inducing atomic coherence between $\ket{g}$ and $\ket{h}$ states.
The waists of the coupling and signal beam in the clouds near field were set to \SI{650}{\micro \meter} and \SI{350}{\micro \meter} respectively. The maximum value of the Rabi frequency for the coupling beam was $\Omega_C = 2 \pi \times \SI{6.9}{\mega \hertz}$. 
We chose the magnetic field gradient $\beta = 2 \pi \times\SI{1.25}{\mega\hertz\per\cm}$, setting the GEM bandwidth to B $= 2 \pi \times \SI{0.9}{\mega \hertz}$ and setting time-bandwidth product to $TB$ = 9, we obtained the memory efficiency $\eta = 34 \%$ . The pulse propagating through the entire length $L = \SI{7}{\milli \m}$ atomic ensemble is delayed by $\tau_{max} = \SI{10}{\micro \s}$. 

\section{Results}
\paragraph*{Sequence - GEM write-in, EIT readout}---
We stored the pulse with the gaussian envelope in the gradient echo memory (GEM), to map the different frequency components of the signal to the specific positions in the atomic cloud as shown in Eq.~\ref{eq:gem}. By changing the frequency of the coupling beam, we can choose the part of the cloud where the pulse is stored.

After the storage time $T_1$, the magnetic gradient is reversed and applied for the time $T_2$ to unwind the GEM storage phase. The stored coherence is read out with the coupling pulse tuned to the two-photon resonance, changing the single photon detuning to $\Delta = \SI{0}{\mega \hertz}$, satisfying the slow-light readout condition. 
The frequency of the signal was varied across the bandwidth of the atomic memory. Due to the frequency-to-position mapping in the GEM, different frequencies acquire different delays according to the length of the propagation distance through the atomic medium. 
If the bandwidth of the stored pulse is substantially smaller than the memory bandwidth, the readouts with different frequencies will be shifted in time due to different delays caused by propagation through the memory. This regime is depicted in the Fig.~\ref{fig:frequency_delay25_3D}\subfig{a}. To each of the results, we fitted a Gaussian envelope as shown with solid lines in the figure. The increase in Gaussian widths stems from considerable dispersion introduced by the EIT. 
However, if the pulse bandwidth is comparable to the memory bandwidth, the light will be stored in the entire cloud.
This way, frequency shifts of the signal will have very little influence on the delays of the readout, because every pulse will have almost identical propagation distance under the slow-light conditions. This regime is depicted in the Fig.~\ref{fig:frequency_delay25_3D}\subfig{b}. The experimental data showing the comparison between delays of the readout in the two regimes are depicted in Fig.~\ref{fig:sim_EIT_freq_timeshifts}\subfig{a,b}. The experiment is showing very good qualitative agreement with the numerical simulation of the light propagation presented in Fig.~\ref{fig:sim_EIT_freq_timeshifts}\subfig{c,d}.

\paragraph*{Sequence - EIT write-in, GEM readout}---
To test the reversibility of the previous protocol, we switched the order of the write-in and readout. 
The signal light is sent to propagate through the atomic memory under the EIT condition, without the presence of a magnetic gradient. To stop the light in the memory, the intensity of the coupling is gradually decreased, slowing the propagation. When the coupling intensity reaches 0, the light is frozen in the atomic ensemble.
To map the position in the cloud to the frequency of the readout, the GEM protocol is performed, i.e., the magnetic gradient is applied for the time $T_1$ and then reversed to unwind the magnetic phase for the time $T_2$. Stored light is then read out from the memory under the GEM conditions with the coupling beam detuned from the single photon resonance by $\Delta = \SI{30}{\mega \hertz}$, mapping the position and width of the stored signal to the frequency. The narrower the pulse was in the time domain during the EIT storage, the narrower it will be in the frequency domain during the GEM readout. Described properties are depicted in the Fig.~\ref{fig:GEM_eit_delay05_3D} \subfig{a} and Fig.~\ref{fig:GEM_eit_delay05_3D} \subfig{b}    
\paragraph*{Verification of reversibility of the EIT-GEM storage}

\begin{figure*}[t]
\centering
\includegraphics[width=2\columnwidth]{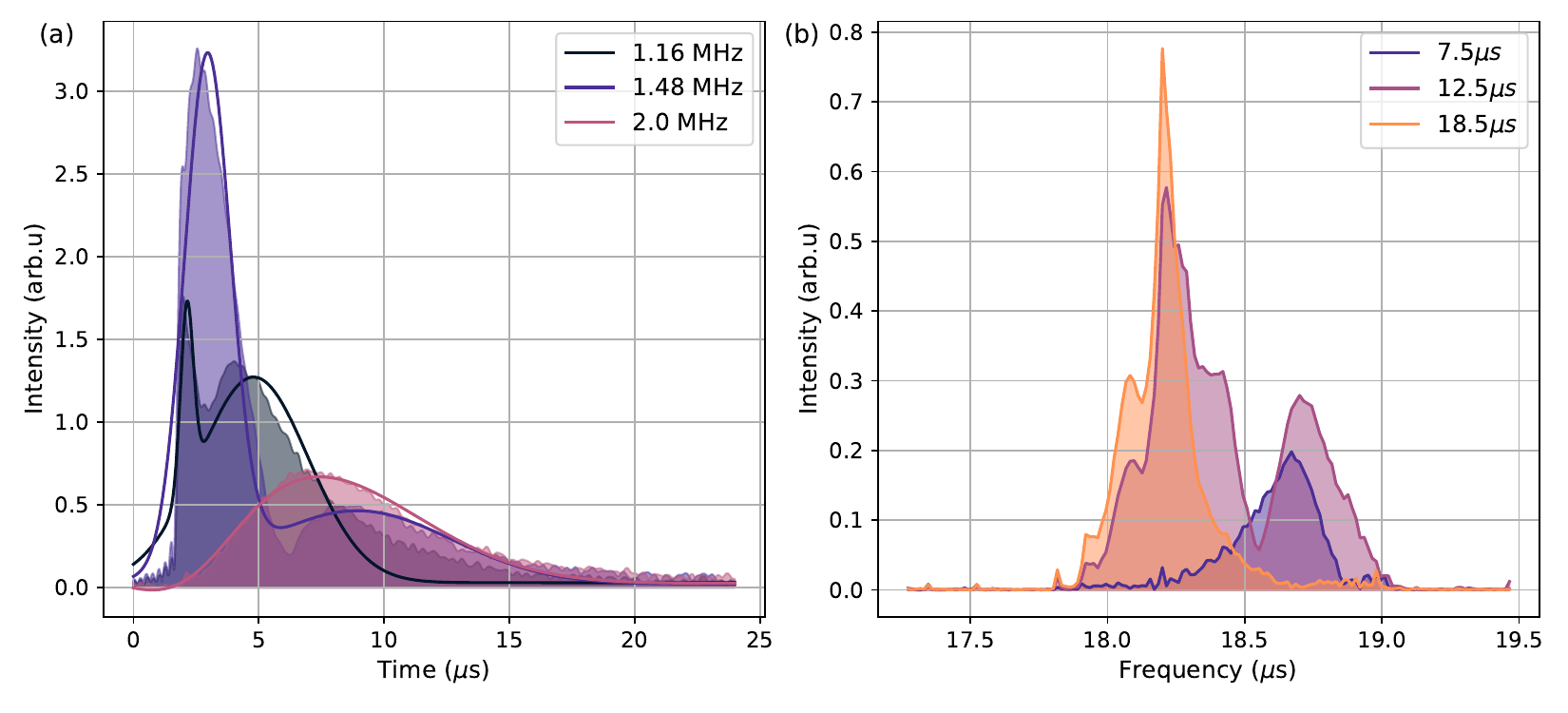}
\caption{\subfig{a} Pulse with 2 different frequencies stored in GEM for different central frequencies of coupling beam and read out in the EIT. The frequencies mapped onto the coherence in the different parts of the cloud are read out with different delays in the EIT due to the difference in the propagation distance. The solid lines show a fitted Gaussian envelope. When one of the frequencies is stored at the beginning of the cloud, it is maximally delayed due to the longest propagation distance. However, when two frequencies are stored in different parts, they are delayed differently, hence two impulses separated in time emerge during the readout.
\subfig{b} Two pulses separated in time stopped in EIT and then readout in GEM regime. Fourier transform of the detected readout signal shows that two frequencies are shifted in frequency according to their position on the cloud. When the impulses are delayed such that both of them are stopped on both ends of the cloud, the GEM readout allows for reading them as two different frequencies. On the contrary, when only one of the impulses is stored, either at the beginning or at the end of the cloud, during readout, we can detect only one of the frequencies, depending on the position where the impulse was stopped on the cloud.  
}
\label{fig:frequency_delay_twopulses_3D}
\end{figure*}

To verify the reversibility of the two approaches and to compare experimental results with the theoretical simulations, we recreated the protocols demonstrated as the simulation in the Fig.~\ref{fig:slow_GEM_schemat}\subfig{e,f}. Therefore, we stored the pulse with two different frequencies, separated by $\delta_{\omega} = \SI{1}{\mega \hertz}$ in the GEM configuration, creating the atomic coherence in the two separated parts of the atomic ensemble. Similarly to before, the magnetic gradient is reversed to unwind the phase of the stored pulse. Then, the signal is read out in the EIT condition. For different central frequencies of the write pulse, the signal is stored in different parts of the cloud, and thus, the frequencies are delayed according to their position on the cloud. This situation is depicted in the Fig.~\ref{fig:frequency_delay_twopulses_3D}\subfig{a}, where in the time domain, two pulses, corresponding to the different position of the coherence in the memory, are delayed differently.
This allows to map the position of the coherence to times of detection.
The second verification is achieved by stopping the two pulses separated by $\delta_{t} = \SI{1}{\micro \s}$. Similarly to a single pulse, two separated pulses were stopped in the atomic ensemble by reducing the intensity of the coupling. Later, the magnetic gradient was applied for the GEM readout protocol. Changing the delays of the input signal results in shifting the central frequency of the readout, and thus, two peaks with variable frequency appear in the signal spectrum. This is confirmed by experimental data shown in Fig.~\ref{fig:frequency_delay_twopulses_3D}\subfig{b}. This way, we can show the mapping of the delays of the impulses to the frequencies detected during the readout.  
Those two presented sequences show the reversibility and compatibility of both regimes, staying in good agreement with numerical simulations. 
Moreover, the combination of those two approaches allows for the reversible time-to-frequency mapping, allowing for the hybrid investigation of the atomic coherence in the ultracold ensemble. 

The readout signal for all experimental realizations was detected using a heterodyne-type measurement, allowing for direct detection of the frequency of the signal. We collected data for 200 experimental sequences, which were then coherently averaged to maintain the phase information. The custom shot noise-limited differential photodiode (DPD) detects the beating between the signal and reference, allowing for phase-sensitive detection. 

\section{Conclusions}
In summary, we have demonstrated two distinct approaches to atomic coherence mapping by combining the GEM and the EIT storage protocols in the $\Lambda$ atomic memory. 
We stored the light in the GEM for different central frequencies of the signal and read it out under the EIT condition. 
By changing the spectral width of the signal beam, we demonstrated the theoretically predicted coherence mapping. 
We have also stored the signal pulse with two frequencies, illustrating the ability to map the spectrum to the time of arrival. 

To test the reversibility of the protocol, we have stopped the light in the EIT and then read it out with the GEM. 
By changing the distance of the propagation of the signal pulse through the memory, we were able to reverse the situation from the first realization of the protocol. 
By storing two separated pulses, we demonstrated the inverted mapping, i.e., mapping of the time of arrival to the frequency. 

The current parameters of the experimental setup (efficiency $\eta = 34\%$, bandwidth $B = 2\pi \times \SI{0.9}{\mega \hertz}$, number of modes $TB = 9 $) enable the proper realization of the presented protocol; however, they can still be improved. 
The main issue in this setup is the trade-off between the storage efficiency and the number of modes stored in the memory.
However, with the same efficiency, longer pulses can be stored by appropriately reducing the memory bandwidth, while keeping the same $TB$ product.  
Increasing the density of the atomic ensemble would allow for better efficiency as well as for the increased delay due to the propagation of the slow light polariton. 
Improving the bandwidth of the memory by either increasing the magnetic field gradient or elongating the ensemble would be especially required for the shorter pulses. 

The protocol allows for reversible conversion between spectral and temporal modes with possible applications in communication systems and metrology, allowing, for example implementation of optical time-of-flight spectroscopy. In communication systems, the information can be encoded as the time-bins which are a crucial resource for quantum information and communication \cite{humphreys_linear_2013}. Moreover, simple mapping between time and frequency domains allows for implementing quantum gates between spectral and temporal modes \cite{Brecht2015,Campbell2012}.  

During the process of writing the manuscript, we were made aware of the work of (Papneja et al.) \cite{Papneja:26}. In contrast to our work, it was mainly focused on the time-frequency Fourier transform, while here our goal was to present the reversible coherence mapping in the hybrid EIT and GEM scheme. In the \cite{Papneja:26} authors also focus only on the storage in GEM and readout in EIT, focusing on the creation of the temporal lens.    

We envisage that the results of this article may pave the way for the tomography of the propagation of Rydberg polaritons and mapping of single Rydberg impurities without the use of a camera. The coherence mapping introduced by storing the Rydberg excitation in the GEM and then reading it out in the EIT would allow to perform the tomography of the propagating polariton without the necessity of utilizing the spatial homodyne detection on the camera \cite{Mazelanik2023}. 

\section*{Data Availability}
Data that supports this study has been deposited at \cite{VJVN4P_2025} (University of Warsaw research data repository).

 \begin{acknowledgments}
 We thank Bartosz Niewelt for insightful discussion and comments regarding the manuscript. The “Quantum Optical Technologies” (FENG.02.01-IP.05-0017/23) project is carried out within the Measure 2.1 International Research Agendas programme of the Foundation for Polish Science, co-financed by the European Union under the European Funds for Smart Economy 2021--2027 (FENG). This research was funded in whole or in part by the National Science Centre, Poland, grant no. 2024/53/B/ST2/04040. Publication co-financed from the state budget funds (Poland), awarded by the Minister of Science under the “Perły Nauki II” program, project No. PN/02/0027/2023, co-financing amount PLN 239,998.00, total project value PLN 239,998.00.
\end{acknowledgments}

\bibliographystyle{apsrev4-2}
\bibliography{refs}
\end{document}